# Thinking is Bad:
## Implications of Human Error Research for Spreadsheet Research and Practice

**Raymond R. Panko,**
College of Business Administration
University of Hawai`i
2404 Maile Way
Honolulu, HI, USA
Ray@Panko.com

**ABSTRACT**

In the spreadsheet error community, both academics and practitioners generally have ignored the rich findings produced by a century of human error research. These findings can suggest ways to reduce errors; we can then test these suggestions empirically. In addition, research on human error seems to suggest that several common prescriptions and expectations for reducing errors are likely to be incorrect. Among the key conclusions from human error research are that thinking is bad, that spreadsheets are not the cause of spreadsheet errors, and that reducing errors is *extremely* difficult.

## 1 INTRODUCTION

In past EuSpRIG conferences, many papers have shown that most spreadsheets contain errors, even after careful development. Most spreadsheets, in fact, have material errors that are unacceptable in the growing realm of compliance laws. Given harsh penalties for non-compliance, we are under considerable pressure to develop good practice recommendations for spreadsheet developers and testers.

If we are to reduce errors, we need to understand errors. Fortunately, human error has been studied for over a century across a number of human cognitive domains, including linguistics, writing, software development and testing, industrial processes, automobile accidents, aircraft accidents, nuclear accidents, and algebra, to name just a few.

The research that does exist is disturbing because it shows that humans are unaware of most of their errors. This "error blindness" leads people to many incorrect beliefs about error rates and about the difficulty of detecting errors. In general, they are overconfident, substantially underestimating their own error rates and overestimating their ability to reduce and detect errors. This "illusion of control" also leads them to hold incorrect beliefs about spreadsheet errors, such as a belief that most errors are due to spreadsheet technology or to sloppiness rather than being due primarily to inherent human error.

## 1.2 Which Error Research?

Most people who are involved with spreadsheets have a good grasp of spreadsheet error research. However, spreadsheet error research is only an insignificant dot in the total picture of human error research.

Given this wealth of research in many fields, focusing exclusively or even primarily on spreadsheet error research makes little sense. Such a narrow focus would bring the not invented here syndrome to a new level of silliness. Unfortunately, this has been done to an uncomfortable degree in past spreadsheet research and writing.

To jump ahead slightly, we should focus especially on research on programming errors and software testing/inspection because experiments and field inspections of operational spreadsheets have shown that software errors and spreadsheet errors are very similar in error rates, detection rates, and types of errors.

## 1.3 Convergent Validation

Every research methodology has both strengths and weaknesses. Consequently, scientists try to provide convergent validation, in which they measure the same construct using different methodologies. If most or all methods give comparable outcomes, then it is possible to have confidence in the results [Campbell & Stanley, 1963].

Figure 2 shows how convergent validity can work in spreadsheet error research. First, in the spreadsheet realm, there are surveys, experiments, and inspections ("audits") of operational spreadsheets. However, there also are general human error studies, including software error studies.

*Figure 1: Convergent Validity*

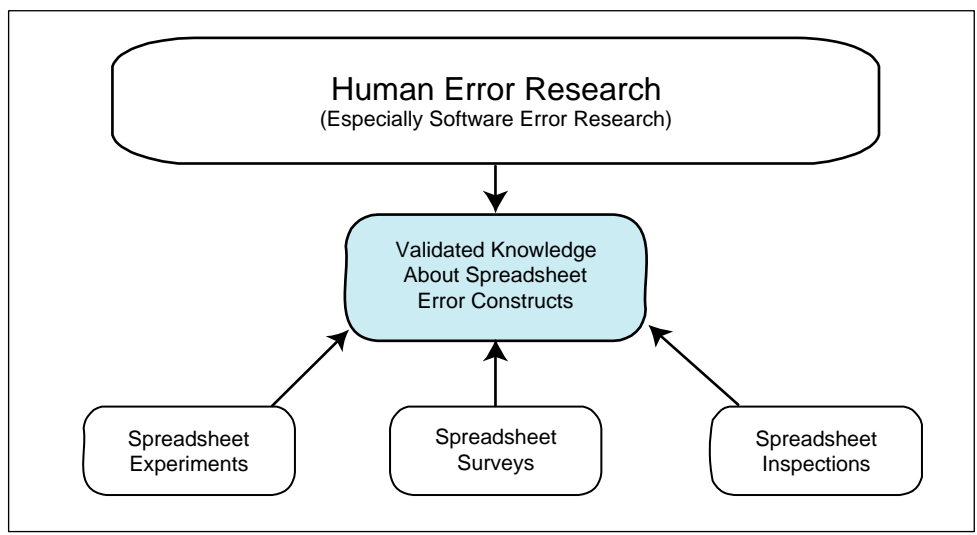

Although it is true that all spreadsheet research studies have high convergent validity [Panko, 2007b], these studies must have high convergent validity with human error studies in other cognitive domains. As importantly, human error studies in other cognitive domains are likely to suggest practices that we can study to reduce errors in spreadsheets and are also likely to call into question some common spreadsheet good practice prescriptions that run counter to research in other areas.

## 2 ERROR FREQUENCY DATA

At the most basic level, human error research in many domains has measured the frequency of errors and error detection rates.

### 2.1 Error Rates by Level

Some of the best human error research has been done in writing. Research on human writing, for instance, has shown that writing is extremely mentally demanding. When a person is typing or writing, they are also concerned about the grammar of the entire sentence as well as what he or she has already said and what he or she still plans to stay to complete the story or argument [Flower & Hayes, 1980; Hayes & Flower, 1980].

Figure 3 shows error rates by level. For mechanical actions, such as typing characters or words, human accuracy is 99.5% to 99.8%. At the level of complex thoughts, such as sentences, accuracy falls to 95% to 98%. At the level of a document, accuracy is 0%, in the sense that any document of nontrivial length will contain grammatical and spelling errors [Panko, 2007a].

*Figure 2: Error Rates by Level*

|  | Writing | Programming | Spreadsheet | Accuracy |
|---|---|---|---|---|
| Mechanical Action | Typing Spelling | Typing Parentheses | Typing Pointing | 99.5%-99.8% |
| Complex Thought (C) | Sentence Grammar, Meaning | Statement, Line of Code (LOC) | Formula | 95%-98% |
| Product ($C^N$) | Document | Program Module | Spreadsheet Module | 0% |

In programming, these three levels correspond to typing actions, statements (lines of code), and programs (or preferably program modules). The accuracy percentages shown in the last column also hold for programming [Panko, 2007a].

Spreadsheet research has also shown the error rates shown in the last column [Panko, 2007b]. This is good because if spreadsheet research had gotten different results, the spreadsheet research would certainly be wrong.

### 2.2 Probability of a Spreadsheet Error

Most people are comfortable with the error rates in Figure 3 for typing errors and pointing errors and even for complex thoughts (formulas); they have a more difficult time accepting that nearly all spreadsheets contain errors. One reason for doubt is that people are not good at understanding how errors multiply. Given an error rate of 5% at the formula level, many people believe that there is only about a 15% or 20% probability of an error in the entire spreadsheet.

Actually, if the error rate per formula is C, then the probability of an error in the spreadsheet as a whole is approximately $C^N$, where N is the number of root (noncopied) formulas in the entire spreadsheet. Note in Figure 4 that even for small spreadsheets with only 10 root formulas, an accuracy rate of 95% to 98% at the formula level translates into a substantial probability of an error (18% to 40%). As soon as the number of root formulas rises to 100 (which actually is quite small for corporate spreadsheets), errors are almost 100% certain. Even if formula error rates are an order of magnitude smaller, large spreadsheets will almost certainly have errors.

*Figure 3: The Probability of a Spreadsheet Error*

| Root Formulas / Formula Accuracy | 95% | 98% | 99% | 99.50% |
|---|---|---|---|---|
| 10 | 40.1% | 18.3% | 9.6% | 4.9% |
| 50 | 92.3% | 63.6% | 39.5% | 22.2% |
| 100 | 99.4% | 86.7% | 63.4% | 39.4% |
| 500 | 100.0% | 100.0% | 99.3% | 91.8% |
| 1,000 | 100.0% | 100.0% | 100.0% | 99.3% |
| 10,000 | 100.0% | 100.0% | 100.0% | 100.0% |

**2.3 Error Detection Rates Are Lower**

So far, we have seen that people are 95% to 98% accurate when they create formulas. However, when people try to *detect* errors, their accuracy rates are much lower.

- In proofreading, for instance, spelling errors are the easiest to detect. However, even for nonword errors, in which the mistake does not create a new word (for example, ther for there), the detection rate is only about 85% to 95%. For word errors, in which the mistake gives another dictionary word (love for live), detection rates are only about 70%. For grammatical errors, detection rates are far lower [Panko, 2007a].

- In software inspections, in turn, individual inspectors only find 40% to 60% of all errors when they work alone, and this falls off if inspection is rushed or too lengthy. Group inspection can increase this yield to 70% to 80% of all errors [Panko, 2007a].

- Several experiments have been done on spreadsheet inspection [Galletta, et al., 1993, 1997; Panko, 1999]. They have shown that individual inspectors find 50% to 60% of seeded errors. This is consistent with proofreading success rates, software inspection rates and other detection rates in other human cognitive activities [Panko, 2007a].

All of these studies used 100% inspection, in which the person looked at each and every word, statement, or formula.

**3. ERROR TAXONOMIES**

It seems reasonable to say that to control errors, you must understand them. A number of error taxonomies have been put forward to illuminate the differences between different types of errors so that different countermeasures can be put into place for different errors if appropriate.

**3.1 Criteria for an Error Taxonomy**

Which error taxonomy is the best? There is no answer to that question because different taxonomies have different strengths and weaknesses. However, we should ask several questions to assess any taxonomy.

- First, we should ask whether the taxonomy can be applied reliably in inspections of finished spreadsheets or whether they can only be applied during observation of the actual development. If a method is developed for observation, its application to inspections of finished spreadsheets must be carefully scrutinized and validated.

- Second, do, different categories should have different error occurrence rates, different error detection rates, require different avoidance or detection methods, or have some other operational importance. Distinctions that do not lead to differential actions need justification.

- Third, from a research viewpoint, awe must validate any measurement methodology for construct validity before it is used in a publication. This is a broad topic, but one aspect of construct validity is straightforward to measure and is a litmus test for publications. This is inter-rater reliability. Unless different people can get the same results when they apply a methodology to an artifact, the methodology lacks reliability. The most common way to measure reliability is to have multiple raters apply the methodology to a set of artifacts and then assess their consistency with Cronbach's [1971] alpha or some newer reliability measure.

**3.2 Norman and Reason**

One of the most widely-used taxonomies is the mistakes/slips/lapses taxonomy created by Norman [1984] and Reason [Reason & Mycielska, 1982]. The basic distinction is between mistakes and slips or lapses. Mistakes are errors in intention. If we intend to do something that is inherently incorrect, this is a mistake. In contrast, lapses and slips are errors in executing an intention. If we have the right intention but the wrong execution, this is a slip or lapse, depending upon whether the mistake is do to a memory or sensory-motor failure.

The attractive thing about the Norman and Reason taxonomy is that it can be used in root cause analysis. However, it is designed to explain thinking and execution, so it is used most naturally in observational research. Construct validity analysis would have to be done to see if the taxonomy can be applied reliably to finished spreadsheet artifacts.

**3.3 Rasmussen and Jenson**

Another error taxonomy that was created for observational studies was the Rasmussen and Jenson [1974] taxonomy, which was developed for studies of experienced professionals with strong sets of rules for what to do in various circumstances. Rasmussen found three types of errors.

- Skill-based errors are basically slips or lapses in highly skilled behaviour.

- Mistakes are divided into rule-based or knowledge-based errors depending on whether the person has a well-established rule (first see if it is plugged in) or whether the person must use general knowledge to solve the problem. The distinction between rule-based and knowledge-based errors is important because when the subject does not have a rule and must resort to general knowledge, execution is likely to be slow with a high potential for errors.

However, the Rasmussen and Jenson taxonomy was created for detailed observational studies. Applying it to finished artifacts is questionable because you normally must be able to hear someone's protocol analysis verbalizations to know whether the person is applying a rule or general knowledge. Also, unless the person has considerable expertise, they will not have the well-developed rule base needed to apply the taxonomy.

**3.4 Panko**

Panko and Halverson [1996, 1997] created a taxonomy with several distinctions:
- First, it distinguishes between quantitative errors (which give the wrong number immediately) and qualitative errors (which are likely to lead to wrong numbers later.)

- Second, based on Allwood's [1984] work in mathematics, the taxonomy distinguishes between mechanical errors, logic errors in creating formulas, and omission errors (leaving something out of a model). This trichotomy of quantitative errors was found to correspond different error occurrence rates [Panko and Halverson, 1997] and, in a limited test, different error detection rates [Panko, 1999].

- Third, for logic errors, it distinguishes between Eureka errors [Lorge and Solomon, 1955], which are easily proven, and Cassandra errors, which are difficult to prove to be errors even if they are detected. Cassandra errors cause particular problems in error-seeking.

Figure 5 shows the author's revised error taxonomy. Based on research in writing [Flower & Hayes, 1980; Hayes & Flower, 1980], which suggests three simultaneous phases—mechanical typing or writing, sentence construction, and overall plan—this paper replaces the mechanical / logic / omission distinction with a different trio. Mechanical errors are called slips or lapses to conform with the Norman and Reason distinctions. Logic errors are simply called formula errors to distinguish between logic errors in individual formulas and logic errors in the overall model design. Most significantly, omission errors are replaced with the more general concept of modeling errors, which are errors in the overall model (omission are only one type of modeling errors).

*Figure 4: Panko Error Taxonomy*

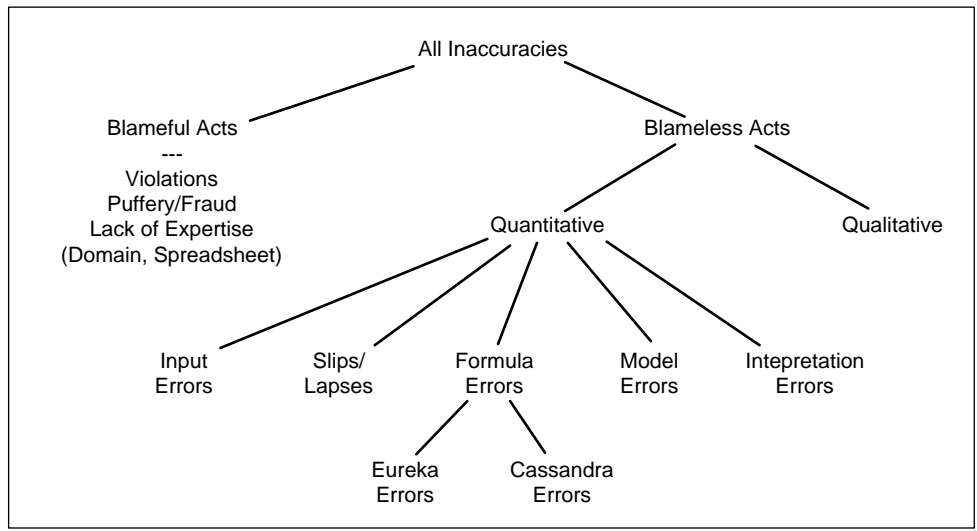

The justification for the slip/lapse-formula-model distinction is show in Figure 6. Here, detection rates are shown to be different, based on various spreadsheet development and testing studies. Although the numbers are rough, they show that the triad appears to be useful in distinguishing error conditions.

*Figure 5: Error Rates for Different Types of Errors*

| Error Type | Spreadsheet Example | Detection Rate |
|---|---|---|
| Slip / Lapse | Typing<br>Pointing | 85% to 99% |
| Formula | Formula Algorithm Incorrect or Expressed Incorrectly | 40%-60% |
| Modeling | Incorrect Requirements<br>Omissions<br>Many others | ~0% |

The model also adds input and output errors to the traditional slip/lapse-formula-modelling triad. This was left out of the original Panko/Halverson taxonomy because their research did not look at problems with dirty data for input or output problems such as misinterpretations. Obviously, these need to be added.

Most errors are blameless. Reason [1990] has noted that most human errors are due to the quirky ways in which our memory and cognition work, not because we are sloppy or because we use a spreadsheet. Humans evolved to have reasonable accuracy at reasonable cost and speed. Unfortunately, when bottom-line values in spreadsheets require hundreds or thousands of calculations, human cognition's inherent limitations are insufficient.

However, *some* inaccuracies truly are blameful. Figure 5 lists three examples of blameful activities.

- Violations are willful departures from required practices. Violations would be departures from the company's standards in spreadsheet development and testing.

- Puffery is making the results look better than they are by making questionable assumptions or choosing modeling decisions that strengthen your case. A certain amount of puffery is likely to be inevitable, but there is a grey zone beyond which puffery is culpable. More extremely, fraud consists of misrepresenting information or withholding information in order to convince another person to take actions against their economic interest or the interests of the firm. Fraud is a criminal offense.

- A final blameful action shown in the figure (although there undoubtedly are many more), is building spreadsheets despite a lack of expertise. Everyone stretches their spreadsheet ability or domain ability to some extent, but to build spreadsheets when the developer lacks sufficient expertise is culpable behavior.

### 3.5 Perspective

Overall, errors are like multiple poisons, each of which is fatal. Different errors may require different countermeasures for reducing their number and for doing testing.

In particular, one cannot simply examine the "hard parts," such as long formulas. Certainly, long formulas are more likely to be wrong than short formulas. However, there are likely to be far more short formulas than long formulas, so it is likely that most errors occur in the "easy parts." "Cherry picking" in testing or inspection is a poor idea.

### 4 ERROR RESEARCH: SUGGESTIONS

A good theory should suggest useful practices. We will look first at some possible suggestions for reducing errors based on error research. We will then look at several areas where current research and recommended practice may be questionable based on error research.

### 4.1 Software Errors and Spreadsheet Errors are Particularly Similar

One general suggestion is to look at the software literature on development errors, testing, and inspection. As noted earlier, the research to date indicates that spreadsheet error rates and detection rates are very similar to those found in software research. So are types of errors. This may surprise many people because software statements can be very complex. However, most software lines of code are very simple, and compilers or interpreters catch most errors in the placement of parentheses and other weirdness in programming.

However, if there really are parallel error and detection rates in spreadsheets and programming, this should be a sobering idea. Software researchers have shown that error is extremely persistent. No "silver bullet" has been found to reduce errors to negligible levels. In particular, experience in testing and inspection has found that roughly a third of all time must be spent in these to activities to have reasonable error detection rates [Panko, 2006b].

### 4.2 Reduce the Memory Load in Modeling

Linguistics has shown that a good way to *cause* errors is to increase someone's memory load. As discussed earlier, writers must simultaneously consider typing, their current sentence, and the overall flow. The same helps in spreadsheeting. This suggests that such actions as modularization and outlining the overall module before building individual parts could reduce memory loads and so could reduce errors.

### 4.3 Do Not Rush

A third way to induce errors in linguistics is to ask someone to work very fast. As speed increases, error rates increase. This suggests, consistent with software research, that when we seek errors through inspection we must work slowly.

Probably, we should slow way down when we reach complex parts of the model during development and inspection. When drivers approach traffic hazards where they need to slow down, they usually slow down only slightly—not to a degree consistent with the danger in the hazard. We probably need to guard against this tendency in development and inspection.

### 4.4 Error Correction during Development

Allwood's [1984] study of students solving mathematical problems used protocol analysis to look in detail at their development process. He discovered that they corrected many errors as they went along. Sometimes, they stopped and did error hunting when they were merely suspicious that "something was wrong." In addition to this general error suspicious, they sometimes had error hypotheses in which they through that a particular problem had occurred.

In many cases, the subject went back to check part of his or her work without having any sense of a problem. It was simply a precaution. These "standard check" episodes often found problems that error hypothesis episodes and error suspicion episodes failed to find.

One way to apply standard checks in spreadsheet development this is to check every formula after entering it. One way is to double-click on a formula cell or hit F2. Both show the formula and in color show the precedent cells. (This provides more information than the arrows functions on the reviewing toolbar.) It is important to hit escape after checking a formula to avoid changing it after using double-clicking or F2.

### 4.5 Use Short Formulas

A common good practice recommendation is to keep formulas short. However, Raffensperger [2003] has suggested the opposite—making formulas long instead of having a string of shorter formulas in order to reduce the length of links to precedent cells. This should increase understandability and should reduce link error in referring to precedent cells because the cells should be nearby.

The evidence to date seems to support keeping formulas short. In proofreading for words of different lengths, detection fell off extremely rapidly with word length [Panko, 1999]. In a spreadsheet inspection study, Panko [1999] found lower detection in longer formulas. However, the frequency of pointing errors to precedent cells was not studied. Direct research is still needed.

**5 ERROR RESEARCH: CHALLENGES**

In addition to suggesting new areas for spreadsheet research and practices, error research can cause us to question some generally accepted prescriptions for reducing spreadsheet errors.

**5.1 Safe and Effective Prescriptions**

In medicine, drugs must be proven both safe and effective before they are sold. If it is not effective, why bother? If it is not reasonably safe, the drug may do more harm than good. We must do the same thing for spreadsheet error research prescriptions. We need to conduct controlled experiments to test prescriptions for safety and effectiveness. Simple consensus is not enough. You cannot vote on truth. Fundamentally, our general lack of awareness of errors makes human perceptions beliefs to be suspicious.

**5.2 Inspection**

A number of studies have been done on spreadsheet inspection to look for errors. Here, the relevant human error research is software inspection research, although proofreading research should also be useful. Fortunately, the developer of software code inspection, Fagan [1976] added the tenet that results should be recorded and analyzed. This has led to a number of controlled studies in the real world. Based on those experiences, software inspections have evolved to include the following main points.

- Understand the requirements and the model logic before you inspect. This is critical for identifying a failure to meet requirements, omissions, and other model errors. Usually, code inspection begins with a meeting in which the author explains the requirements, flow of the model, and specific sections of code.

- Do not rush. Limit inspections to about two hours, and limit inspections to about 200 lines of code. When the rate of inspection goes up, error detection falls.

- Use team inspection. As noted earlier, inspection research in software and in spreadsheets has consistently shown that individuals catch only 50% to 60% of errors. Team inspection in spreadsheets can raise the detection rate to 80%. The gain comes almost entirely from the types of errors that individuals have the most difficult time detecting [Panko, 1999].

**5.3 Use of an Assumptions Section: IPO versus Flow**

The area that probably requires the biggest "reality check" is the desirability of or lack of desirability of using an assumptions section that contains all of the input data. The processing section of the model then will contain formulas with links back to the assumptions section.

One reason for suggesting the use of an assumptions section is that it allows the quick and accurate detection of all input cells. Second, software and statistical programs have traditionally separated programs files from data files (although object-oriented programming no longer separates all code from data).

Counting against assumptions sections, many formulas in the processing section will have to refer to distance cells in the assumption section. In an exploratory pilot study that was not published, the author found that when subjects had to point to precedent cells that were off the current screen, their error rates went up substantially. In addition, if the assumptions section is put on a different worksheet in an Excel workbook than the processing sections, neither F2/double-click nor precedent errors will give useful information to help check for errors.

Comprehension may also be reduced. When a data cell is looked at out of its processing context, it may be difficult to understand. Note that in all common forms, such as the infamous U.S. 1040 income tax form, the information is presented as a flow with data and formulas interspersed in a logical way.

Overall, we probably need to conduct a good deal of research on the suitability of input-processing-output (IPO) worksheet designs and flow worksheet designs.

**5.4 Static Checking Programs**

One way to reduce errors in word processing is to use the spelling checker function. In software development, this is called static testing. However, spell checkers are only good for finding nonword errors. They are likely to miss all world errors. People can do much better than this.

Grammar checkers are even more problematic. Galletta et al. [2005] found that students who used grammar checkers had *poorer* grammar in essays than those who did not. Analyzing the data, they found that students who used grammar checkers often took incorrect suggestions. Static testing programs do not really identify errors. They really say, "Look here" and perhaps given a correct or incorrect reason why.

Static inspection programs for spreadsheets generally have two functions. The first is to show the logic flow in the program or program module at a very high level, so that the inspector can divine the flow. This can be extremely helpful, but is it as effective as doing a preparation session in code inspection where this is done more formally?

Second, the static inspection program should identify errors. Unfortunately, there are certain spreadsheet errors for which static testing programs are likely to be ineffective. For instance, if the spreadsheet developer omits a variable, expecting a static inspection program to find it seems unreasonable. In addition, while some pointing errors will give indications of errors, others do not. Finally, if the developer has the wrong algorithm for a formula or expresses it incorrectly in terms of order of precedents, parentheses, or other matters, can the static checker find most or all of these errors?

Doing a prescan with a static inspection program is almost certainly a good idea. However, it is important not to expect too much from this. Given the number of static testing vendors and the enthusiasm of their claims, the goal of proving prescanning safe and effective is especially important.

We must understand what errors static testing programs identify and do not identify in order to have an idea of how much work the individual inspector must do after prescanning. Is the human part just a "mop up" operation, or is the end of prescanning simply where the real work begins?

## 5. CONCLUSION

The basic argument in this paper is a request that academics and professionals in spreadsheet error research draw on the vast human error literature to inform their research, work, and general thinking. Almost all of the results that we have seen in spreadsheet experiments or field inspections have been predictable in light of prior findings in human error research. In addition, this paper has shown a few samples of how human error research can suggest new research and practice directions and of how at least some aspects of research and recommended practice that we may need to challenge.

It is fundamentally important to understand that the problem is not spreadsheets. Asking what is wrong with spreadsheets has some value, but it is fundamentally the wrong issue. The real issue is that thinking is bad. Of course, thinking has some benefits, and we are accurate 96% to 98% of the time when we do think. But when we build spreadsheets with thousands (or even dozens) of root formulas, the issue is not whether there is an error but how many errors there are and how serious they are.

The second fundamental thing to understand from human error research is that making large spreadsheets error free is theoretically and practically impossible and that even reducing errors by 80% to 90% is extremely difficult and will require spending about 30% of project time on testing and inspection. Trusting "eyeballing" and "checking for reasonableness" to reduce errors to a reasonable level flies in the face of a century of human error research as is a testament to the profound human tendency to overestimate their ability to control their environment even when they cannot.

The third fundamental thing to understand is that replacing spreadsheets with packages does not eliminate errors and many not even reduce them. The problem, again, is that thinking is bad. Unless there are no human decisions or design actions in using a package, there will be thinking, and therefore there will be errors. In addition, packages often do only part of what must be done, requiring people to use spreadsheets for many risky calculations. A good package should reduce errors, but this is speculation that can and should be tested.

## REFERNCES